\documentclass{article}
\usepackage{bm} % fixes boldsymbol
\usepackage{amsmath} % Comment out for acmart
\usepackage{amssymb} % Comment out for acmart
\usepackage{amsthm} % Comment out for acmart
\usepackage[boxed]{algorithm}
\usepackage[noend]{algpseudocode}
\usepackage{pifont} % xmark, cmark
\usepackage{braket}

\algnewcommand\algorithmicinput{\textbf{Input:}}
\algnewcommand\Input{\item[\algorithmicinput]}
\algnewcommand\algorithmicoutput{\textbf{Output:}}
\algnewcommand\Output{\item[\algorithmicoutput]}
\algnewcommand\algorithmicalgorithm{\textbf{Algorithm:}}
\algnewcommand\Algorithm{\item[\algorithmicalgorithm]}

\newtheorem{theorem}{Theorem}[section]

\newtheorem{proposition}{Proposition}[section]

\newtheorem{assumption}{Assumption}[section]

\usepackage[utf8]{inputenc}
\usepackage{listings}
\usepackage{enumerate}
\usepackage{graphicx}
\usepackage{xcolor}
\usepackage[backref=page]{hyperref}
\usepackage{setspace}
\usepackage{subcaption}
\usepackage{fullpage}
\hypersetup{
    colorlinks,
    linkcolor={blue},
    citecolor={blue},
    urlcolor={blue}
}
\RequirePackage[T1]{fontenc}  % Comment out for acmart
\RequirePackage[tt=false, type1=true]{libertine} % Comment out for acmart 
\RequirePackage[varqu]{zi4} % Comment out for acmart
\RequirePackage[libertine]{newtxmath} % Comment out for acmart

\title{I/O complexity and pebble games with partial computations}

\author{Aleksandros Sobczyk 
\\IBM Research and ETH Zurich
\\Zurich, Switzerland\footnote{Work performed exclusively while at IBM Research and ETH Zurich. Current affiliation: Huawei Technologies Switzerland AG.}}
\date{}

\usepackage{graphicx}
\usepackage{xcolor}

\begin{document}

\maketitle

\begin{abstract}
Optimizing data movements during program executions is essential for achieving high performance in modern computing systems. This has been classically modeled with the Red-Blue Pebble Game and its variants. In  existing models, it is typically assumed that the number of red pebbles, i.e., the size of the fast memory, is larger than the maximum in-degree in the computational directed acyclic graph (DAG).
Graphs that do not satisfy this constraint need to be first transformed appropriately, which is not a trivial task for general graphs. In this work we propose a Pebble Game variant to model DAGs with arbitrary in-degrees, by allowing \emph{partial computations}. In the new model, we show that it is NP-complete to decide whether there exists an optimal pebbling strategy with cost $k$, even for single-level DAGs and when only two words fit in the fast memory. Approximation algorithms for special cases are also outlined.
\end{abstract}

\section{Introduction}
Input/Output (I/O) complexity has evolved the past decades as a way to analyze algorithms and programs in terms of minimizing data movements between the memory levels in a system, rather than the number of arithmetic or logic operations that are executed. It was first introduced by
Hong and Kung \cite{jia1981complexity}, together with the so-called \textit{Red-Blue Pebble Game} (RBPG), followed by Aggrawal and Vitter who formally defined the so-called  \textit{I/O model of computation} \cite{aggarwal1988input}.  
In this model, there is a small fast memory (cache) that can fit $M$ \text{words} of arbitrary type, e.g., real numbers, or bit-strings, and an arbitrarily large slow memory. There is also a Central Processing Unit (CPU), that can execute operations between words that reside in the fast memory.  

In the RBPG, computer programs are represented by computational Directed Acyclic Graphs (DAGs). Similar to standard arithmetic or Boolean circuits, nodes with zero in-degree are called \textit{inputs}, and nodes with zero out-degree are  \textit{outputs}.  Intermediate nodes represent computations involving their predecessors. 
There are $M$ red pebbles, representing the fast memory, and arbitrarily many blue pebbles, denoting the main memory. The execution of a program is modeled by moving pebbles over the DAG, based on a set of rules (which are summarized in Table \ref{table:rules_of_pebble_games}). 
The goal is to pebble the entire DAG with the minimum number of pebble moves (memory transfers).

The main goal of this work is to stress a limitation of the RBPG, namely, the following Assumption \ref{assumption:max_indegree}, which is present (explicitly or implicitly) in existing model variants.
\begin{assumption}
    \label{assumption:max_indegree}
    In the RBPG, the maximum in-degree of any node in the DAG is at most $M-1$.
\end{assumption}
Many real-world applications rely on computational kernels that can be described by low-depth arithmetic circuits with large (non-constant) in-degrees. Sparse matrix operations are such an example, which lie in the core of atomistic simulations \cite{hutter2014cp2k}, data analytics \cite{mahoney2011randomized,woodruff2014sketching}, optimization \cite{boyd2004convex,gill2021numerical}, iterative methods  \cite{saad2003iterative,saad2011numerical}, graph algorithms \cite{demaine2018fine,kepner2011graph}, and Large Language Models (LLMs) \cite{vaswani2017attention,zaheer2020big}.
In order to model such DAGs with large in-degrees in the RBPG, the underlying graph must be first transformed to an equivalent one that satisfies Assumption \ref{assumption:max_indegree}. In the context of I/O complexity,
an ideal transformation should preserve the I/O cost of the optimal RBPG solution in the transformed graph.  
\begin{figure}[htb]
    \centering
    \subcaptionbox{\label{fig:cdag_addition_a}}[.14\textwidth]
    {
        \includegraphics[width=\linewidth]{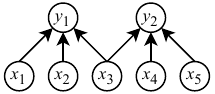}
    }
    \subcaptionbox{\label{fig:cdag_addition_b}}[.16\textwidth]
    {
        \includegraphics[width=\linewidth]{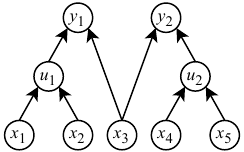}
    }
    \subcaptionbox{\label{fig:cdag_addition_c}}[.16\textwidth]
    {
        \includegraphics[width=\linewidth]{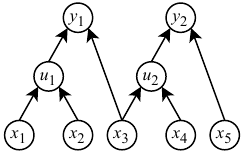}
    }
    \caption{ (a) Example DAG with max in-degree three. (b) and (c) are two equivalent transformations with max in-degree two.}
    \label{fig:different_cdags_example}
\end{figure}

It is argued that such transformations are not trivial for general DAGs (as one might expect, since other relevant graph transformations are known to be NP-complete  \cite{aho1976optimal,aho1972optimization,bruno1976code}). To give some intuition, consider the example in Figure \ref{fig:different_cdags_example} (a), where we have five inputs, $x_1,x_2,x_3,x_4,x_5$, and we want to compute two sums: $y_1=x_1+x_2+x_3$ and $y_2=x_3+x_4+x_5$. Assume that we want to optimize the I/O complexity for this DAG in the original RBPG with $M=3$ red pebbles. To solve the problem, we need to transform the DAG to an equivalent one with maximum in-degree two, for example, by replacing every node with large in-degree with a balanced tree, as shown in
Figures \ref{fig:different_cdags_example} (b) and (c).
The optimal RBPG solution in the DAG of Figure \ref{fig:different_cdags_example} (c) has cost $7$, which is achieved by executing computations left-to-right. However, in the equivalent DAG of Figure \ref{fig:different_cdags_example} (b), there is no solution with cost less than $8$. 
By taking more elaborate examples, it can be shown that the cost of a naive transformation can increase the I/O cost significantly, beyond a constant factor.\footnote{Proposition 4.7 in the recent work of \cite{papp2025impact}  implies that a naive transformation can increase the cost from $O(1)$ to $\Theta(n)$.} 
Based on these observations, it is natural to ask whether we can design a model to incorporate I/O complexity analysis for graphs with arbitrary degrees, without enforcing prior transformations, and what is the complexity of finding optimal solutions in such a model. 

\paragraph{Contributions.}
The main contributions are the following.
\begin{enumerate}[(i)]
    \item We describe a pebble game which enables the I/O complexity analysis of computational DAGs with large in-degrees. This is achieved by allowing the computation of partial results, which can be stored in the main memory and used again in later steps.
    \item In this new model, it is proved that it is NP-complete to decide whether the cost of an optimal solution (pebbling strategy) is equal to $k$. Interestingly, the proof holds even for single-level DAGs, and even when only two words fit in the fast memory. 
    \item We describe approximation algorithms for single-level DAGs, which can important operations such as sparse matrix products. For $M=2$, we report a $\tfrac{21}{8}$ approximation ratio, which can be improved to $\tfrac{8}{7}$ if the machine can perform a LOAD and STORE operation simultaneously in a single step.  
\end{enumerate}  

\paragraph{Prior work.}
Several works have studied the hardness of pebble games in the literature, even before the RBPG was proposed. \cite{sethi1975complete} proved that pebble games for register allocation problems are NP-complete, and a few years later the so-called Standard Pebble Game was shown to be PSPACE-complete \cite{gilbert1980pebbling,paul1978time}. 
More recent results proved that such games are not only hard to solve, but also to approximate  \cite{demaine2017inapproximability}. 
The hardness of the original RBPG was established by Liu and Demaine \cite{demaine2018red,liu2017red}, who showed that finding a pebbling strategy with optimal cost is PSPACE-complete. The authors also showed that it is NP-complete if zero-cost pebble deletions are forbidden (this is often called the \textit{NODEL} variant). Papp and Wattenhofer \cite{papp2020hardness} showed that some other important variants are also NP-complete, including some inapproximability results. An extension of the model of \cite{jia1981complexity} over memory hierarchies was described in \cite{savage1995extending}. \cite{demaine2018fine} studied several sparse graph problems from the fine-grained complexity perspective in the I/O model.
Recently, \cite{bohnlein2024red} proposed a rigorous  extension for multiple processor machines, including several hardness and approximability results.

Regarding algorithms for finding optimal pebbling strategies, previous works have reported optimal algorithms for important structured DAGs, such as matrix multiplications and Fast-Fourier Transforms 
\cite{jia1981complexity,kwasniewski2019red,toledo1997improving}. For sparse/irregular DAGs, several works have analyzed elegant algorithms, even reporting  worst-case optimality guarantees \cite{bender2010optimal,gleinig2022optimal,greiner2012sparse,greiner2010complexity,pagh2014input}, albeit without approximation guarantees for general (non-worst-case) instances.
Approximation algorithms in the context of I/O complexity have received little attention in general. The only other works that we are aware of to report actual approximation guarantees (for other game variants/problems) are \cite{bohnlein2024red,carpenter2016brief}. 
To our knowledge, this is the first work to report constant-factor approximations, even for single-level DAGs.

\section{Pebble game and hardness}\label{sec:mod_rbpg_}
In this section we describe the proposed model. 
We are given some \textit{inputs}, and a set of \textit{associative operations}, which receive an arbitrary number of inputs (or \textit{operands}) and they produce one output. 
\begin{table*}[bht]
\small
\caption{Definition of the original RBPG and the proposed model.}
\label{table:rules_of_pebble_games}
\renewcommand{\arraystretch}{1.3}
\begin{tabular}{r p{6.8cm} p{6.5cm}}
\hline
  &
  Proposed model& 
  RBPG 
  \\\hline\hline
\textbf{Pebbles}
  & 
There are $M$ red and $M$ yellow pebbles. At most $M$ pebbles (any color) are present on the DAG at a time.
&
There are $M$ red pebbles and arbitrarily many blue pebbles available.
\\
\textbf{Start}
  & 
None of the nodes have any pebbles on them.
& 
Only the inputs have pebbles (blue color).
\\
\textbf{End}
  & 
All edges are deleted and the outputs have no pebbles.
& 
All outputs have blue pebbles on them.
\\\hline
\textbf{Rule (Cost)} & & \\
LOAD  (1) & A red pebble is placed on a node with no pebbles. & A red pebble is placed on a node with a blue pebble.
\\
REMOVE (0) & 
A red pebble is removed from a node. & 
A red pebble is removed from a node. 
\\
COMPUTE (0)
&
If $u,v$ have pebbles (of any color) and $u$ is a leaf, then edge $(u,v)$ is deleted. The pebble on $v$ turns yellow. 
& If $u$ and $v$ have red pebbles and they are both predecessors of $w$, a red pebble can be placed on $w$.$^{(\dagger)}$ 
\\
STORE 
(1)
& 
A yellow pebble is switched to red. 
& Place a blue pebble on a node with a red pebble. 
\\\hline
\multicolumn{3}{l}{$^{(\dagger)}$\footnotesize{ In the ONESHOT RBPG variant, only the first COMPUTE on each node has zero cost (re-computations have infinite cost).}}
\end{tabular}
\end{table*}
An operation with $k$ inputs can be represented as a single-level DAG: there is a target vertex (the output), $k$ input vertices, and an edge from each input to the output. 
A sequence of operations, with a corresponding DAG $G=(V,E)$, represents an algorithm. 
The inputs of the DAG have no incoming edges, and the outputs have no outgoing edges. Nodes without incoming edges will be also called \emph{leaves}.

To execute an algorithm, we assume a machine that follows the $(M,B)$-I/O model of computation that was described in the introduction (following the definitions of \cite{aggarwal1988input}).
All inputs and (partial) outputs of mathematical operations are assumed to fit in a single \textit{word}. In the $(M,B)$-I/O model, there is a slow memory (main memory) that can fit an arbitrary number of words. The inputs of the algorithm are initially stored in the main memory. There is also a fixed-size fast memory (cache) which can store $M$ words, and a CPU capable of performing computations between two words that are located in the fast memory. 

When an element is not present in fast memory, it has to be retrieved (loaded) from the main memory. 
Such a retrieval always brings a set of $B<M$ words at the same time.
This set is called a \textit{cache-line}, and $B$ is the cache-line size.  We assume that the inputs and the outputs are all stored in consecutive positions in memory and that they are aligned. To simplify the analysis, we make the following Assumption \ref{assumption:static_cache_line}, which allows us focus on the simpler, $(M,1)$-I/O model. It can potentially be removed without affecting the results, but it is not covered here.
\begin{assumption}[Static cache-line]\label{assumption:static_cache_line}
Given a DAG in the $(M,B)$-I/O model, we assume that each node of the graph, including intermediate ones, is assigned a fixed location in the main memory.
\end{assumption}
\begin{proposition}
\label{proposition:mb_to_m1}
Under Assumption \ref{assumption:static_cache_line}, a  DAG  $G=(V,E)$ in the $(M,B)$-I/O model can be uniquely transformed to an equivalent  DAG $G'=(V',E')$ in the $(M,1)$-I/O model in $O(|E|+|V|)$ time.
\label{prop:MB_M1_equivalence}
\end{proposition}
\begin{proof}
Each node $v\in V$ in the original graph is assigned to a unique location in the main memory, that is, it belongs to a single and fixed cache-line. There exist $k=\lceil|V|/B\rceil$ such cache-lines. For each cache-line we create a new node in $V'$. Clearly, $|V'|=k$. This is the complete set of nodes of $G'$. Now, for each edge $(v,w)\in E$, we create a new edge $(v',w')\in E'$, such that $v'$ corresponds to the cache-line that contains $v$ in the original graph, and $w'$ contains $w$, respectively.
\end{proof}

In the proposed model, there are two pebble colors: \textbf{red} and \textbf{yellow}. Whenever a node has a pebble on it (of any color), it means that it is loaded in the cache.
\begin{itemize} 
\item \textbf{Red} pebbles denote that the value of a node in the cache is the same as its contents in the main memory.
\item \textbf{Yellow} pebbles, on the other hand, signify that the contents of a word in the fast memory have been modified by a partial computation. 
\end{itemize}
A partial computation between two nodes $u$ and $v$ is modeled by deleting the corresponding edge $(u,v)$. For this to happen it is required that both $u$ and $v$ have pebbles on then (of any color), and that $u$ is a leaf, i.e., it has no incoming edges. The last requirement ensures that the value of $u$ has already been finalized, i.e., all the  preceding computations have been executed. 

\paragraph{Game rules.} The rules of the game are the following (summarized in Table \ref{table:rules_of_pebble_games}). The game starts with an ``empty board'', i.e. there are no pebbles on the graph, and it ends when all edges have been deleted and the final outputs have no pebbles on them.

In a LOAD operation, a word / cache-line is copied from the main memory to the fast memory. This is an expensive operation, and it is assigned a cost of one.

A REMOVE operation can be executed on a vertex (word or cache-line) whose contents have not been modified by computations (i.e it has a \textbf{red} pebble), and therefore it can be safely deleted from the fast memory with no cost.

In  a COMPUTE operation, a partial result is formed inside the fast memory after a computation. The color of the target pebble is changed to yellow, to indicate that the contents of the word have been modified, and, as already mentioned, the corresponding edge is deleted. 

A STORE operation is used to save the contents of such a modified word (that has a yellow pebble) back to the main memory. The pebble on the vertex switches back to red, which means that it can be safely removed from the fast memory, since its contents are already saved.

A \emph{pebbling strategy} is a sequence of LOAD, REMOVE, COMPUTE, and STORE operations, that solves the game. The total cost is equal to the total number of LOAD (placing red pebbles on empty nodes) and STORE (changing yellow pebbles to red) operations.

\paragraph*{Hardness.} Given the aforementioned rules, we can study the hardness of the game.

\begin{theorem}
\label{theorem:mod_rbpg_p_hardness}
Let $G=(V,E)$ be a computational DAG and assume a machine in the $(M,1)$-I/O model. The following problem is NP-complete: decide whether there exists a sequence of moves from Table \ref{table:rules_of_pebble_games} that solves the pebble game with partial computations with cost at most $k$.
\end{theorem}
\begin{proof}
 First, note that the longest sequence of pebble moves with zero cost that can be achieved has length one. Any other pebble move that follows will have cost one. Therefore, we can execute a polynomial number of steps of a pebbling strategy to verify that its cost is $k$.

We next prove that it is NP-hard.
Let $G=(V,E)$, $|V|=n$ and $|E|=m$, be an undirected graph for which we want to solve Hamiltonian Path problem. We  construct a bipartite graph $G'$ from $G$, with a special structure, and we seek for an optimal cost pebbling strategy for $M=2$ on $G'$. 
The first step is to define some gadgets. For every node $v_i$ in $G$, we construct a gadget $g_i$, which is a directed bipartite graph: it has $n$ inputs (one for each vertex in the $G$), one output, and there is an edge from all the inputs to the output. The inputs of $g_i$ are denoted as $s_{ij}$, $j\in[n]$ and the output is denoted by $t_i$. For simplicity, we will say that we \emph{eliminate} $g_i$, meaning that all of its edges are deleted, and its final output has been stored back in the main memory.

\begin{figure}[H]
    \centering
    \vspace{-2mm}
    \subcaptionbox{Original graph $G$ for which we need to solve the Hamiltonian Path problem.\label{fig:g_original}}[.45\textwidth]
    {
        \includegraphics[width=0.15\textwidth]{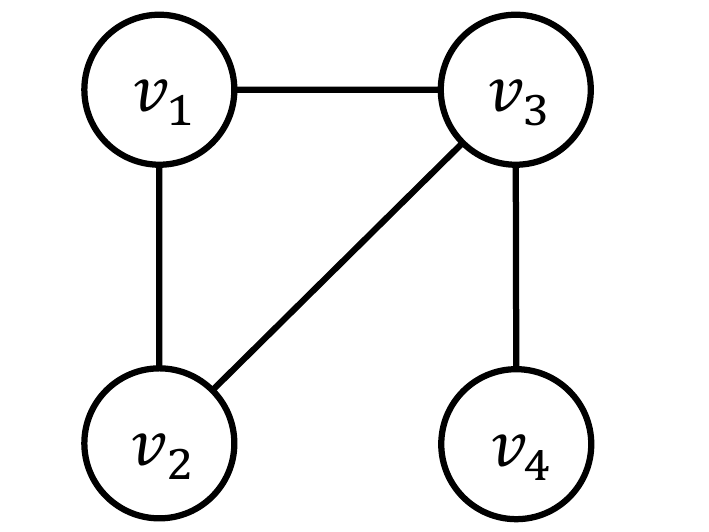}
    }
    \subcaptionbox{Gadgets $g_1,g_2,g_3,g_4$ in $G'$, one for each one of the nodes of $G$.\label{fig:g_gadgets}}[.45\textwidth]
    {
        \includegraphics[width=0.4\textwidth]{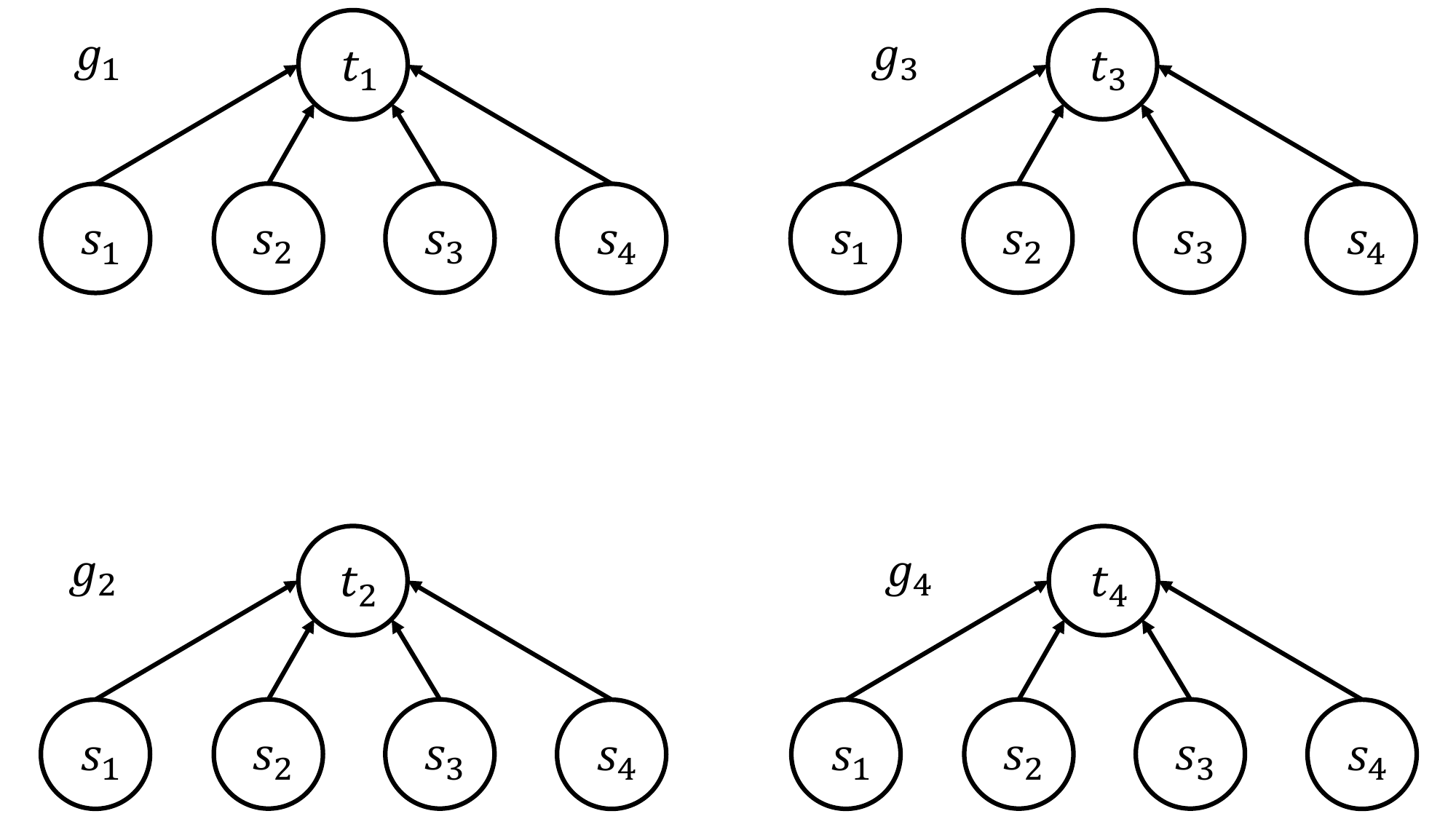}
    }
    \subcaptionbox{Source vertices that must be merged in $G'$ based on the existing edges in $G$.\label{fig:g_gadgets_to_merge}}[.45\textwidth]
    {
        \includegraphics[width=0.4\textwidth]{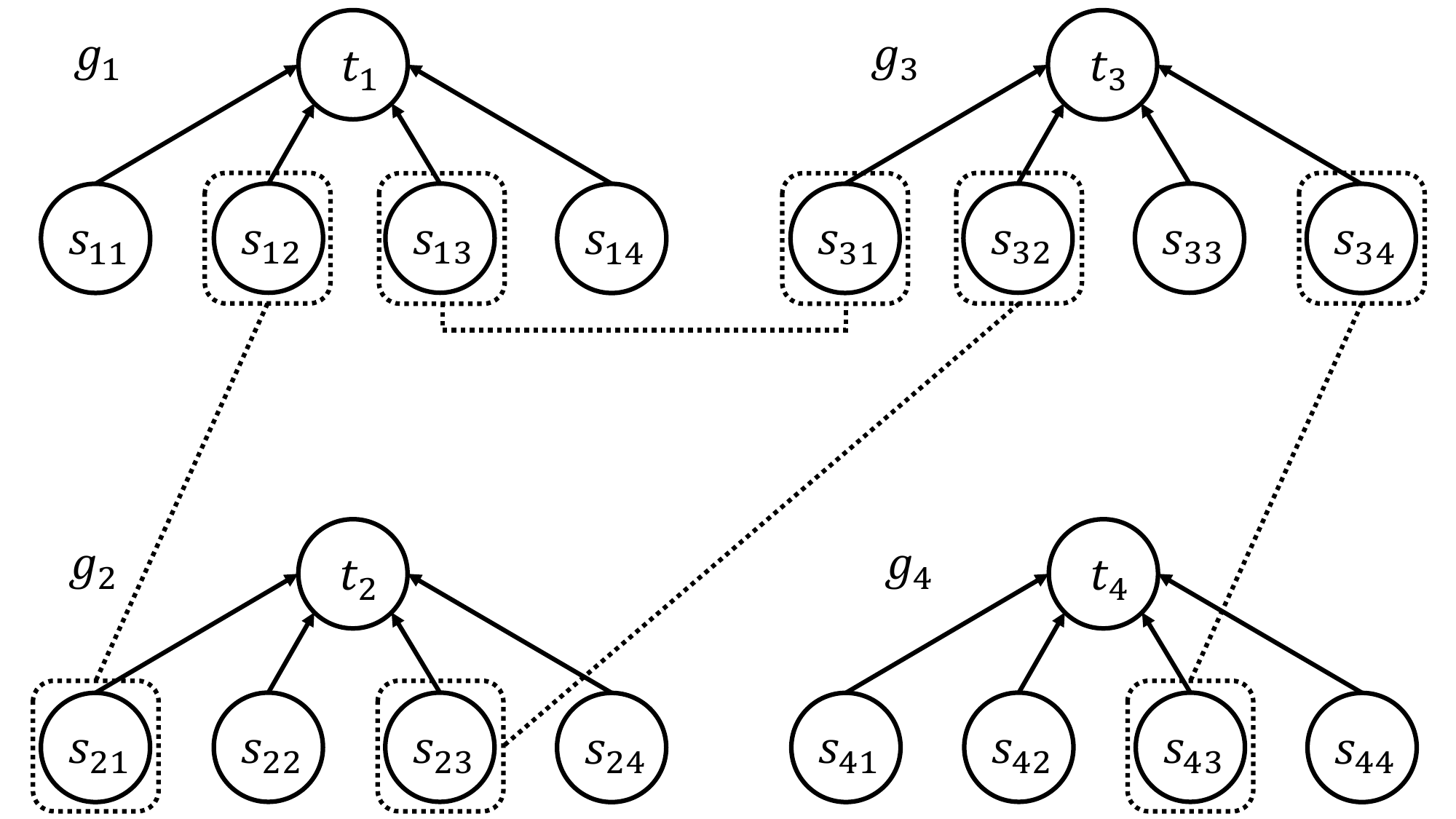}
    }
    \subcaptionbox{Final graph $G'$ after appropriately merging the gadget sources.\label{fig:g_gadgets_merged}}[.45\textwidth]
    {
        \includegraphics[width=0.4\textwidth]{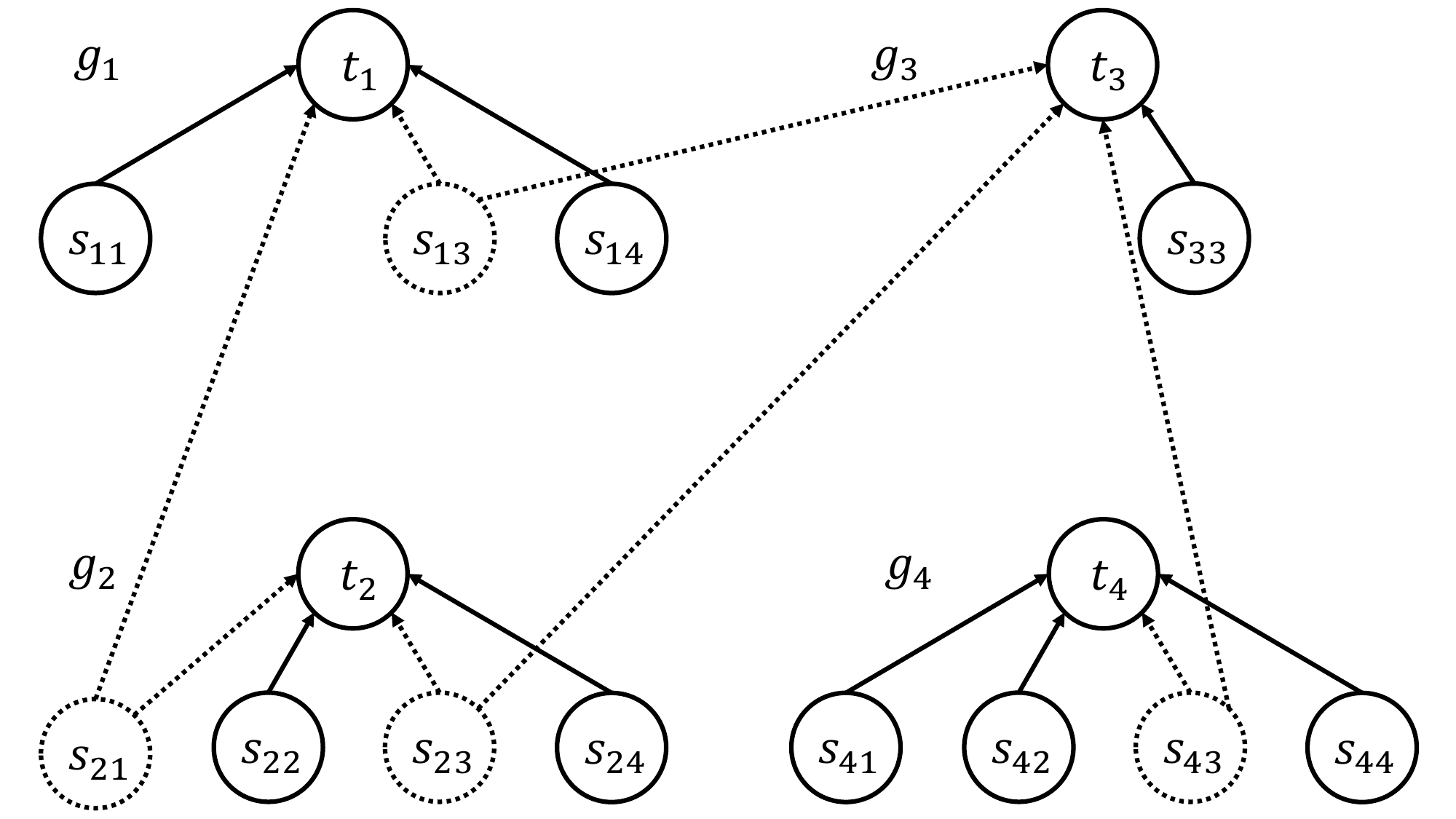}
    }
    \caption{Steps required to create the single-level DAG $G'$ from the graph $G$ in the proof of Thm.\@  \ref{theorem:mod_rbpg_p_hardness}.}
    \label{fig:graph_and_gadgets}
\end{figure}

In Figures \ref{fig:g_original} and \ref{fig:g_gadgets} we illustrate an example graph and its corresponding gadgets.
The elimination of $g_i$ can be done with $n+2$ moves: 
\begin{itemize}
    \item One move to load $t_i$, by placing a red pebble on it;
    \item $n$ moves to sequentially load all $s_{ij}$ and delete all the edges. The pebble of $t_i$ becomes yellow;
    \item One final move to store back the result, by changing the yellow pebble of $t_i$ back to red.
\end{itemize}

Each of the $n$ input nodes $s_{ij}$ of the gadget $g_i$ corresponds to one of the nodes of $G$. The following connection is made. If two nodes $v_i$ and $v_j$ are connected with an edge in $G$, then we merge the $j$-th input of $g_i$, $s_{ij}$, with the $i$-th input of $g_j$, $s_{ji}$ into one new single input with two outgoing edges, one towards $t_i$ and one towards $t_j$. The construction of $G'$ is now complete. Figure \ref{fig:g_gadgets_to_merge} shows the gadgets that need to be merged. Figure \ref{fig:g_gadgets_merged} shows the final bipartite graph $G'$.

Next, consider any pair of vertices $v_i$ and $v_j$ in $G$, and the corresponding gadgets $g_i$ and $g_j$ in $G'$. If there is no edge in $G$ between $v_i$ and $v_j$, then $g_i$ and $g_j$ do not share any input, so we need $2\times (n+1)$ moves to eliminate both $g_i$ and $g_j$. However, if edge $(v_i,v_j)$ exists, then $g_i$ and $g_j$ share an input node. 
We can eliminate the first gadget with $n+2$ moves, visiting $s_{ij}$ last. As $t_i$ has been stored in the main memory (i.e., its pebble was switched back to red), we can safely delete it from the cache at no cost.

The next move is to load $t_j$, the output of the second gadget $g_j$. This deletes the edge $(s_{ij},t_j)$ with a single move. We consecutively load all the remaining $n-1$ sources of $g_j$ in arbitrary order, to eliminate the rest of the edges, with a single move per edge. Finally, we store back the result of $t_j$. This gives a total of
$
    (n+2)+1+(n-1)+1=2n+3
$
moves to eliminate both gadgets, instead of $n+2$.
Note that this is the optimal cost of eliminating two consecutive gadgets: we need at least $2n+4$ moves to eliminate both gadgets if they don't share a source, and at least $2n+3$ moves if they do. Moreover, two gadgets cannot share more than one source, since there are no duplicate edges in $G$. We can finally argue that two gadgets $g_i$ and $g_j$ can be both eliminated with $2n+3$ moves if and only if there is an edge between the corresponding vertices $v_i$ and $v_j$ in $G$.

A Hamiltonian Path in $G$ visits exactly once every vertex, and there exists an edge between all the consecutive vertices that it visits. From the previous paragraph, we can argue that a Hamiltonian Path in $G$ exists if and only if there exists a strategy to eliminate all the edges in $G'$ that costs at most $1\times (n+2)+(n-1)\times (n+1)=n^2+n+1$ moves: $n+2$ moves for the first gadget, and thereafter $n+1$ moves for each one of the subsequent $n-1$ gadgets.
\end{proof}

\paragraph*{Discussion and extensions.}
The proposed model definition has two evident limitations. They do not affect the reported results, and they can be straightforwardly addressed by adjusting the rules appropriately, but we clarify them here for completeness.

The first is that “on-the-fly” intermediate computations are not possible: every \emph{non-input} node must eventually be stored in the main memory, once its final value is completed.
Now, all the DAGs studied in this work consist of a single level, and, therefore,
every non-input node is in fact an output that must be anyway
stored in the main memory. As such, the reported results are not affected.
The second limitation is that, currently, \emph{every} node  in the graph (even intermediate ones) must be loaded from the main memory at least once. This implicitly assumes that all nodes contain an initial value that needs to be taken into account during computations. Important classes of problems do fall in this category, such as (sparse) matrix products of the form $C\gets C+AB$ (``GEMM'' operations), or polynomial evaluations (e.g. via
Horner’s method). Certainly, however, this assumption limits the applicability on general DAGs.

Both these limitations can be  resolved by extending the game rules appropriately, for instance, as described in the recent work of Papp et al \cite{papp2025impact}. The authors provided a more comprehensive definition of the model described here, which also allows a direct comparison with the original RBPG. For example, it was shown that \emph{any} pebbling strategy in the original RBPG can be translated to an equivalent one  in the model of \cite{papp2025impact} with at most the same cost, and that it is NP-hard to decide whether the respective optimal solutions of the two models have equal cost. The interested readers are referred to \cite{papp2025impact} for many additional details and results.

\section{Approximation algorithms
}\label{sec:approximation}
\begin{figure}[htb]
    \centering
    \subcaptionbox{When $v_1,w_1$ are in cache, two pebble moves with a total cost of one are required to delete $(v_2,w_1)$.\label{fig:two_edges_case_a}}[0.3\textwidth]
    {
        \includegraphics[width=0.15\textwidth]{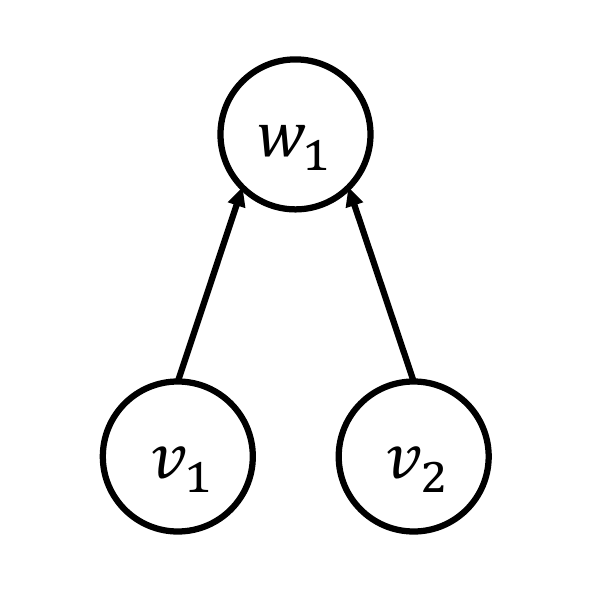}
    }
    \quad
    \subcaptionbox{When $v_1,w_1$ are in cache, three pebble moves are required with a total cost of two to delete $(v_1,w_2)$.\label{fig:two_edges_case_b}}[0.3\textwidth]
    {
        \includegraphics[width=0.15\textwidth]{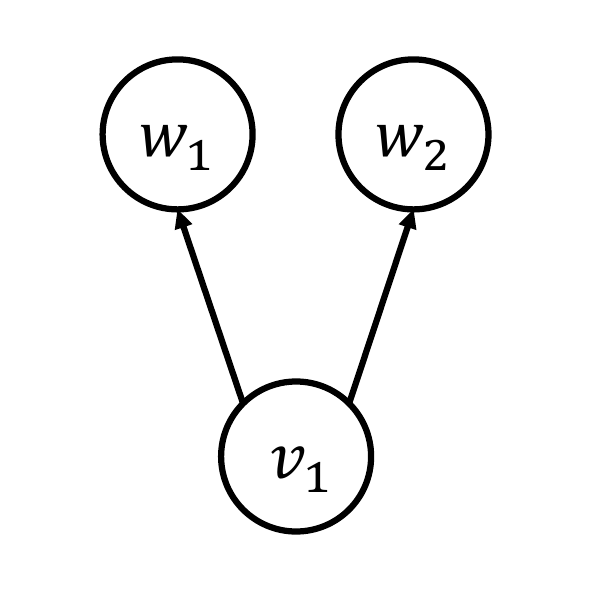}
    }
    \quad
    \subcaptionbox{When $v_1,w_1$ are in cache, five pebble moves are required with a total cost of three to delete $(v_2,w_2)$.\label{fig:two_edges_case_c}}[0.3\textwidth]
    {
        \includegraphics[width=0.15\textwidth]{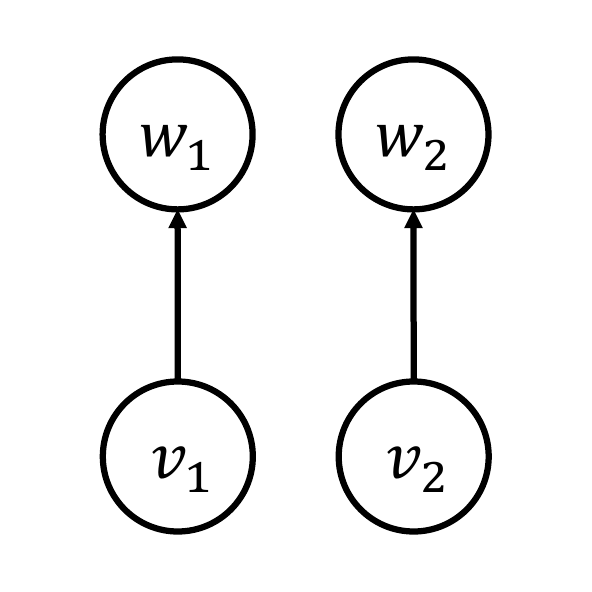}
    }
    \caption{Three different possible configurations of two edges in the $(2,1)$-I/O model, and the associated cost in pebble moves to eliminate the right edge, assuming the left edge was just deleted.}
    \label{fig:two_edges_possible_cases}
\end{figure}

In this section we discuss approximation algorithms.
When $M=2$, we first observe that the deletion of edges simplifies significantly. Assuming that we just deleted edge $e_1=(v_1,w_1)$, and we want to eliminate the edge $e_2=(v_2,w_2)$, the total cost in terms of moves is given by one of the following three cases (illustrated in Figure \ref{fig:two_edges_possible_cases}):
\begin{enumerate}[(a)]
    \item Two pebble moves with total cost one, if they share the same target: REMOVE $v_1$ and then LOAD $v_2$. In this case $w_1=w_2$, which means that we can REMOVE $v_1$ from the cache with no cost, and LOAD $v_2$ with cost one, and then perform the computation at no cost. 
    \item Three moves with total cost two, if they share the same source: STORE $w_1$, REMOVE $w_1$, and then LOAD $w_2$. In this case $v_1=v_2$. We first need to STORE $w_1$ in memory with cost one, then REMOVE it and then LOAD $w_2$ from the main memory with cost one, totaling two moves. 
    \item Five moves with total cost three, if they do not share any vertex: STORE $w_1$, REMOVE $w_1$ and $v_1$, and then LOAD $v_2$ and $w_2$. We first need to STORE $w_1$ back to the main memory with cost $1$, and then we need to LOAD both $v_2$ and $w_2$. 
\end{enumerate}

\begin{proposition}
\label{prop:christofides_approximation}
Let $G=(V,E)$, with $|V|=n$ and $|E|=m$ be a single-level  DAG. Let $OPT$ be the cost of an optimal pebbling strategy in the pebble game with partial computations, with $M=2$. We can find a pebbling strategy with cost at most $\frac{21}{8} OPT$ in polynomial time. If the machine is modified to allow LOAD and STORE to be executed simultaneously as one instruction, then the optimal pebbling strategy can be approximated within a factor of $\frac{8}{7} OPT$ in polynomial time.

\end{proposition}

\begin{proof}
Recall that the game ends when all the edges are deleted, and there are no remaining pebbles present on the graph. When $M=2$, the cost of an optimal solution is in fact the same as the cost of an optimal strategy to just delete the edges of the graph using the same rules, plus one last move to save the output of the last remaining edge. We therefore focus just deleting the edges, based on the rules of Table \ref{table:rules_of_pebble_games}. 

We first construct a complete graph $G'$ with $m$ vertices, one vertex for each edge in $G$. We will then assign weights to the edges of $G'$ based on costs described in observations (a), (b) and (c). For each pair of edges $e_i$ and $e_j$ in $G$, which correspond to the nodes $v_i'$ and $v_j'$, we assign the following weight for the corresponding edge $(v_i',v_j')$ in $G'$:
\begin{itemize}
    \item if $e_i$ and $e_j$ share the same target in $G$, weight$=1$;
    \item if $e_i$ and $e_j$ share the same source in $G$, weight$=2$;
    \item if $e_i$ and $e_j$ do not share any nodes in $G$, weight$=3$.
\end{itemize} 
It is not hard to see that any pebbling strategy in $G$ is a Hamiltonian Path in $G'$, and vice versa.
Therefore, we can find a low-cost pebbling strategy in $G$ by finding a Hamiltonian Path in $G'$ with small cost. The construction of $G'$ is depicted in Figure \ref{fig:mod_rbpg_reduction_to_min_hamiltonian_path}.
\begin{figure*}[htb]
    \centering
    \subcaptionbox{Bipartite graph $G$.\label{fig:graph_g_mod_rbpg_basic}}
    {
        \includegraphics[width=0.45\textwidth]{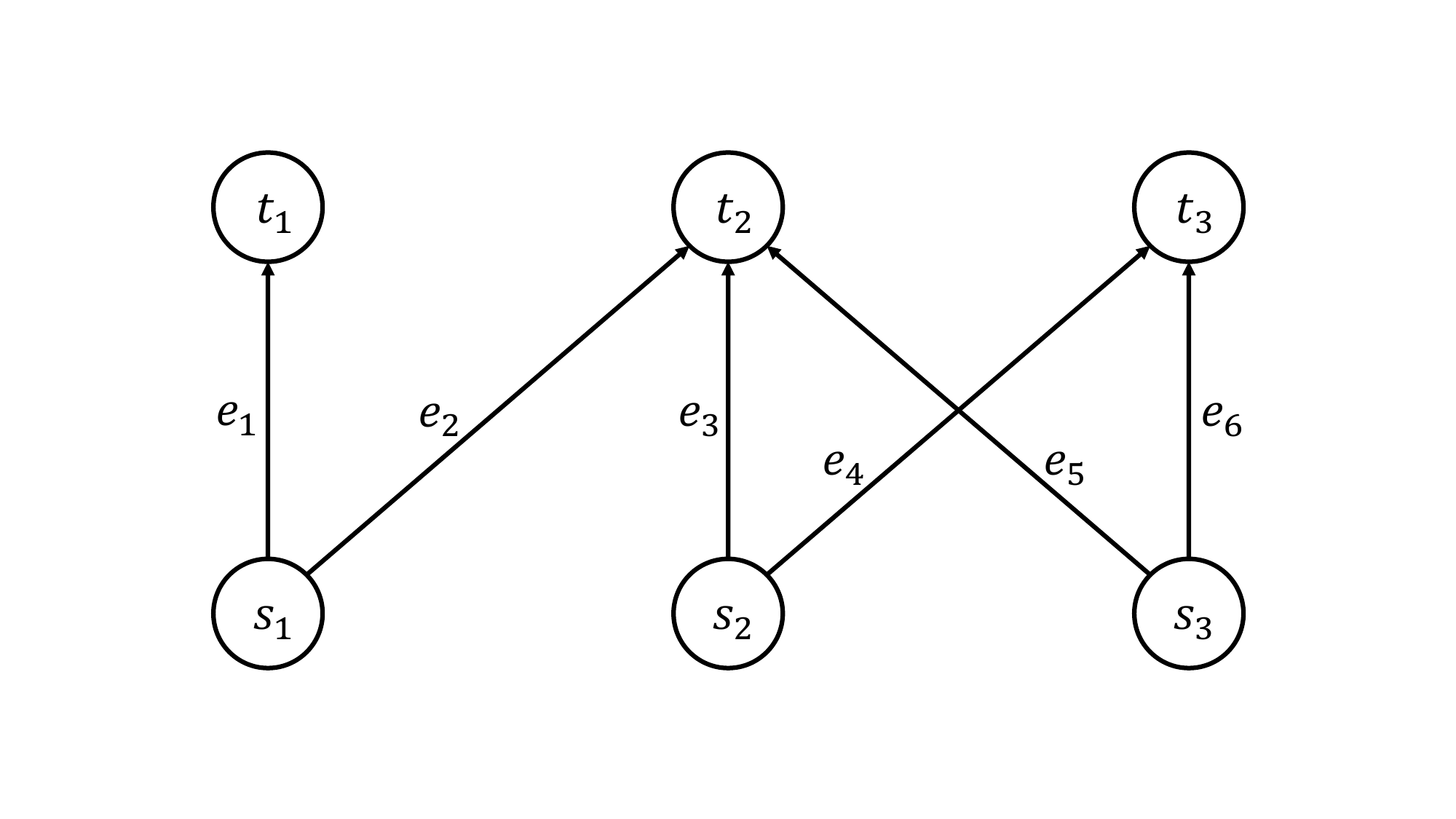}
    }
    \subcaptionbox{Graph $G'$. The dashed edges all have weight equal to $3$.\label{fig:graph_g_mod_rbpg_complete_weighted_line_graph}}
    {
        \includegraphics[width=0.45\textwidth]{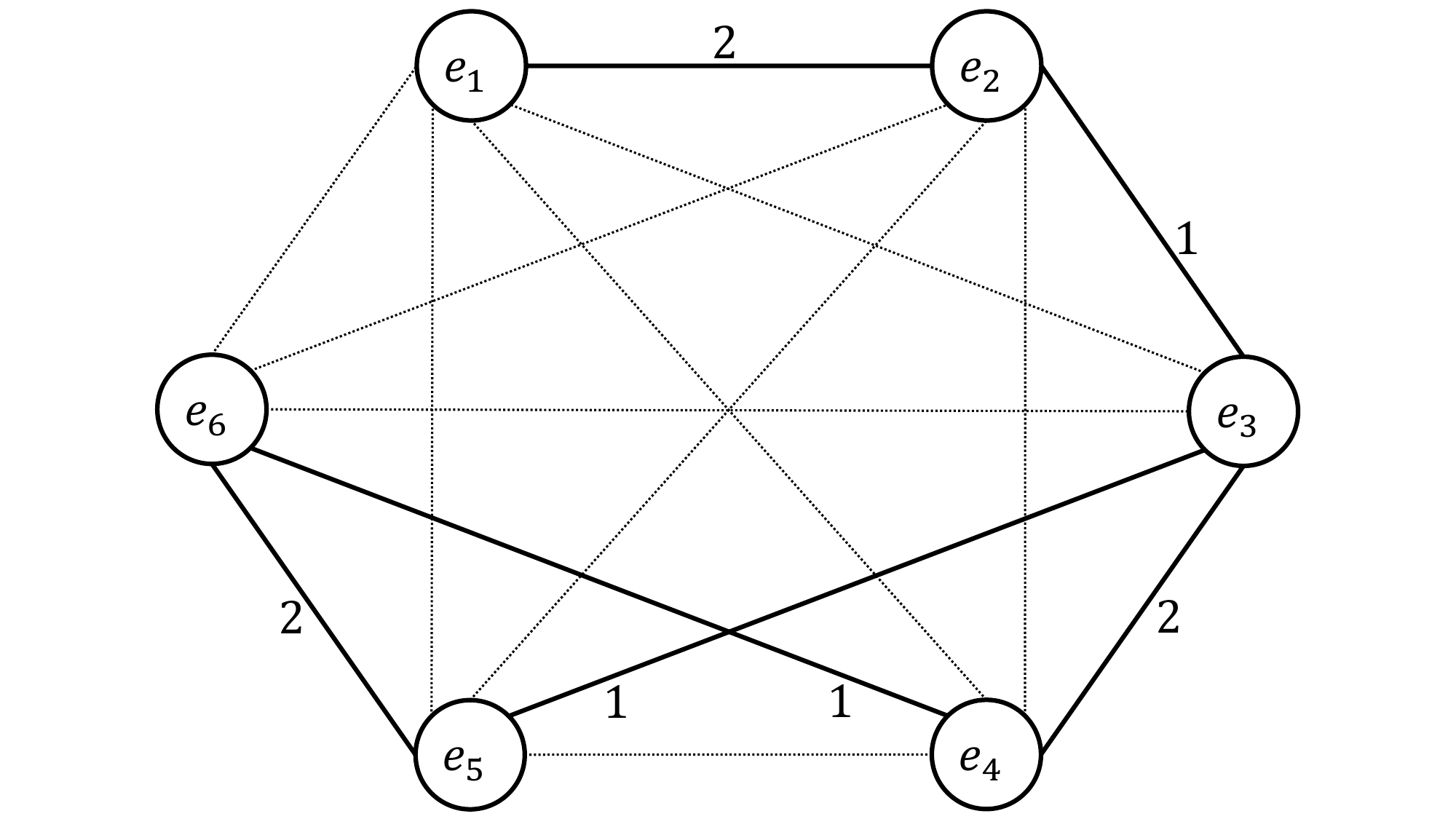}
    }
    \caption{A bipartite graph $G$ and the corresponding graph $G'$ to solve the minimum Hamiltonian Path.}
    \label{fig:mod_rbpg_reduction_to_min_hamiltonian_path}
\end{figure*}
While the edge weights of $G'$ do not satisfy the triangle inequality, their weights are bounded, which allows us to obtain constant factor approximations based on Christofides' algorithm for the metric TSP \cite{christofides1976worst}. The algorithm is as follows. Given a complete undirected weighted Graph $G=(V,E,w)$:
\begin{enumerate}
    \item Create a minimum spanning tree $T$ of $G$.
\item Find a minimum cost perfect matching $M$ on the subgraph induced by the odd-degree vertices in $T$.
\item Merge $T$ and $M$ to form an Euler graph $F$, and find an Euler tour on $F$.
\item Follow the tour, shortcutting duplicate vertices, to create a path $P$, and return $P$.
\end{enumerate}

The rest follows from standard analysis of Christofides' algorithm, 
but we state it for completeness. We have that the edge weights of $G'$ are in $\{1,2,3\}$. Starting from the end, we have that $
    cost(P)\leq \tfrac{3}{2}cost(F), $
since, in the worst-case, each shortcut replaces two edges with weight at least one each, with one edge with cost at most three. 
Next, the cost of $F$ is equal to the cost of the matching $M$ plus the cost of the MST $T$, i.e. $cost(F)=cost(M)+cost(T)$. 
The optimal solution is a path, whose cost is greater than or equal to that of any MST on $G'$. Therefore, $OPT\geq cost(T)$. 
It remains to get an upper bound for $cost(M)$. Let $O$ be the set of vertices with odd degree in $T$. Assume that we follow the optimal Hamiltonian Path $P^*$, and we only keep the vertices of $O$, by short-cutting vertices that do not belong to $O$. This gives a tour $H$, where $cost(H)\leq \frac{3}{2}cost(P^*)$ holds from the same shortcut argument as before. 

We now split the edges of $H$ in two sets $H_1$ and $H_2$, in a round robin: the first edge in $H$ goes to $H_1$, the second edge of $H$ to $H_2$, the third to $H_1$, etc. Both of these sets form a perfect matching on $O$. Since $
    cost(H_1)+cost(H_2)=cost(H),$
we have that either $H_1$ or $H_2$ have a cost at most $cost(H)/2$, which in turn is bounded by $
    cost(H)/2\leq \tfrac{3}{4}cost(C)=\tfrac{3}{4}OPT.$
Therefore, since $M$ is a minimum cost perfect matching, it holds that $
    cost(M)\leq \min\{
        cost(H_1),cost(H_2)
    \}
    \leq 
    \tfrac{3}{4} OPT.$
Finally, we have that $cost(F)=cost(T)+cost(M)\leq \frac{7}{4} OPT$. This gives the first result, i.e., $cost(P)\leq \frac{3}{2}cost(F)\leq \frac{21}{8}OPT$.

For the second result, if we modify the rules to allow LOAD and STORE to be executed simultaneously as one instruction (which is also more realistic for existing architectures), we can further refine the analysis.
By modifying the gadgets in the proof of Theorem \ref{theorem:mod_rbpg_p_hardness}, it can be shown that this variant is also NP-complete.
However, it is possible to obtain better approximation factors for this variant. Indeed, if we perform the same analysis as in Proposition \ref{prop:christofides_approximation}, we can observe that, in this case, the graph $G'$ has edge weights equal to $1$, if the corresponding edges share any endpoint in $G$, and $2$ if not.  This way, with slight modifications we can reduce the problem to the well-studied $(1,2)$-TSP \cite{adamaszek2018new,berman2006,papadimitriou1993traveling}. 
The first work to study this problem depth is \cite{papadimitriou1993traveling} which gave an $7/6$-approximation algorithm. The best known approximation ratio is $8/7$ due to \cite{berman2006}. In recent work, \cite{adamaszek2018new} improved the runtime of this algorithm to $O(n^3)$ and also give an algorithm with $7/6$-approximation ratio which runs in $O(n^{2.5})$.
\end{proof}

\section{Conclusion}\label{sec:conclusion}
This work addresses a limitation of existing variants of the Red Blue Pebble Game, namely, the standard assumption that the in-degrees of the corresponding computational DAGs need to be smaller than the size of the fast memory. To that end, we proposed a  Pebble Game which allows partial computations. It was shown that deciding the cost of an optimal I/O strategy in the proposed model is NP-complete, even for the special case when the corresponding DAG has one level, and only two words fit in the fast memory. We also outlined approximation algorithms for a couple of special cases, by reductions to TSP instances.

% \bibliographystyle{plain}
% \bibliography{main} 

\end{document}